\documentclass{emulateapj}
\slugcomment{Accepted in ApJ}

\begin{document}

\title{ Stellar Velocity Dispersion of the Leo A Dwarf Galaxy }

\author{Warren R.\ Brown,
	Margaret J.\ Geller,
	Scott J.\ Kenyon,
	Michael J.\ Kurtz}

\affil{Smithsonian Astrophysical Observatory, 60 Garden St, Cambridge, MA 02138}
\email{wbrown@cfa.harvard.edu}

\shorttitle{ Stellar Velocity Dispersion of Leo A }
\shortauthors{Brown et al.}

\begin{abstract}

	We measure the first stellar velocity dispersion of the Leo A dwarf galaxy,
$\sigma = 9.3 \pm 1.3$ km s$^{-1}$.  We derive the velocity dispersion from the
radial velocities of ten young B supergiants and two H{\sc ii} regions in the
central region of Leo A.  We estimate a projected mass of $8\pm2.7\times10^7
~M_{\sun}$ within a radius of $2\arcmin$, and a mass to light ratio of at least
$20\pm6 ~M_{\sun}/L_{\sun}$.  These results imply Leo A is at least $\sim$80\% dark
matter by mass.

\end{abstract}


\section{INTRODUCTION}

	The Leo A dwarf galaxy was discovered by \citet{zwicky42} and is one of the
most remote galaxies in the Local Group.  Leo A is gas rich, with an H {\sc i}
velocity dispersion of 3.5 to 9 km s$^{-1}$ and with no observed rotation
\citep{allsopp78, lo93, young96}.  Leo A is also extremely metal poor, with an
abundance of $12+\log{\rm O/H} = 7.3$ to 7.4 measured from H {\sc ii} regions
\citep{skillman89, vanzee06}.

	Photometric studies of Leo A reveal both a red and blue plume of stars in
its color-magnitude diagram indicating recent star formation \citep{demers84,
sandage86, tolstoy96}.  {\it Hubble Space Telescope} observations have resolved the
stellar population of Leo A, which shows evidence for numerous epochs of star
formation spanning billions of years \citep{tolstoy98, schulte02, cole07} as well as
an old stellar ``halo'' \citep{vansevicius04}.  RR Lyrae variables confirm the
presence of an $\sim11$ Gyr old population, and place Leo A at a distance of $800
\pm 40$ kpc \citep{dolphin02}.  Recently, \citet{brown06b} reported the first
spectroscopy of stars in Leo A:  two B supergiants stars observed serendipitously as
part of their hypervelocity star survey.  The B supergiants provide spectroscopic
proof of star formation as recently as $\sim$30 Myr ago in Leo A.

	Inspired by the B supergiant observations, we have obtained spectroscopy for
ten additional blue-plume objects in Leo A.  There is no a-priori reason to expect
that Leo A's steller and H {\sc i} gas velocity dispersions are identical.  
Detailed H {\sc i} maps show velocity structure, which suggests that the gas may be
affected by cooling or may not yet be relaxed \citep{young96}.  Our observations
allow us to measure the stellar velocity dispersion, and thus estimate the mass of
Leo A's dark matter halo.  In \S 2 we discuss our target selection, observations,
and stellar radial velocity determinations.  In \S 3 we present the resulting
velocity dispersion and mass-to-light ratio of Leo A.  We conclude in \S 4.

\section{DATA}

\subsection{Target Selection}

\begin{figure}		
 \plotone{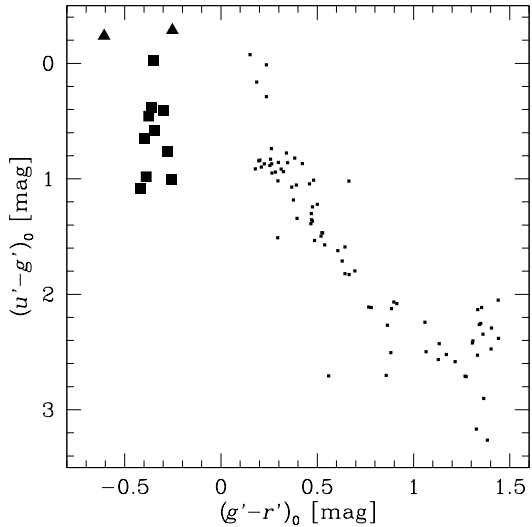}
 \caption{ \label{fig:ugr}
	Color-color diagram of every star in SDSS with $g'<21$ and within $9\arcmin$
of Leo A (centered at $9^{\rm h} 59^{\rm m} 23\fs92$ $+30\arcdeg 44\arcmin
47\farcs69$ J2000).  We target the twelve blue plume candidates with $(g'-r')_0<0$.  
We identify ten B supergiants ({\it solid squares}) and two H {\sc ii} regions ({\it
solid triangles}). }
 \end{figure}

\begin{figure}		
 \plotone{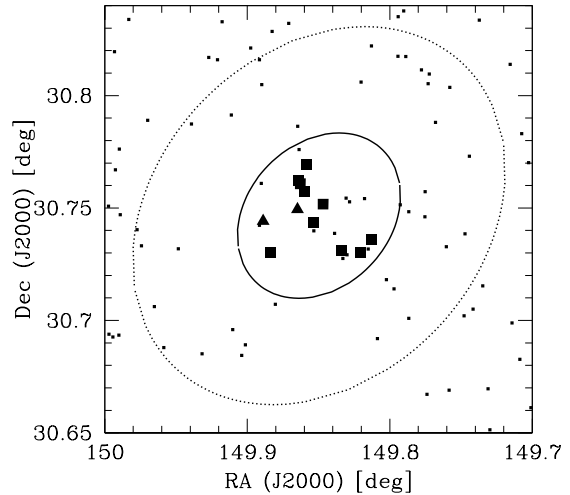}
 \caption{ \label{fig:radec}
	Location of objects in Figure \ref{fig:ugr}, where the symbols are the same 
as before.  For reference, the solid ellipse marks Leo A's Holmberg radius
$a=3\farcm5$ \citep{mateo98} and the dotted ellipse marks the extent of Leo A's
stellar ``halo'' \citep{vansevicius04} and H {\sc i} gas \citep{young96}.}
 \end{figure}

	We use Sloan Digital Sky Survey \citep[SDSS,][]{adelman07} photometry to
select candidate Leo A blue plume stars by color.  
	We illustrate our target selection in Figure \ref{fig:ugr}, a color-color
diagram of every star in SDSS Data Release 5 with $g'<21$ and within $9\arcmin$ of
Leo A (see also Figure \ref{fig:radec}).  We compute de-reddened colors using
extinction values obtained from \citet{schlegel98}; the adopted extinction values
are $E(u'-g')=0.029$ and $E(g'-r')=0.022$.
	Objects with $(g'-r')_0<0$ and $(u'-g')_0<1.1$ are objects in the blue
plume.  The blue plume can contain massive main sequence stars, blue supergiant
stars, and blue-loop stars \citep[e.g.][]{schulte02}.  We target the 12 blue plume
objects with $g'<21$ (solid squares and triangles).

	Objects with $(g'-r')_0>0$ in Figure \ref{fig:ugr} have colors consistent
with foreground stars, ranging from F-type stars at the main sequence turn-off
$(g'-r')_0\sim0.2$ to late M dwarfs $(g'-r')_0>1$.  Stars with $(g'-r')_0\sim1.4$
may include some asymptotic giant branch stars in Leo A.

	Figure \ref{fig:radec} plots the position of every star in Figure
\ref{fig:ugr}.  For reference, the ellipses follow Leo A's observed H {\sc i}
profile, with center $9^{\rm h} 59^{\rm m} 23\fs92$ $+30\arcdeg 44\arcmin
47\farcs69$ (J2000), semiminor to semimajor axis ratio 0.6, and position angle
$104\arcdeg$ \citep{young96}.  The solid ellipse marks Leo A's Holmberg radius
$a=3\farcm5$ \citep{mateo98}, and the dotted ellipse with $a=8\farcm0$ marks the
extent of Leo A's H {\sc i} gas \citep{young96} and stellar ``halo''
\citep{vansevicius04}.  All twelve blue plume candidates are located within
$2\arcmin$ of the center of Leo A; probable foreground objects are distributed more
uniformly across the field.

\subsection{Observations}

	We obtained spectroscopy of the twelve blue plume objects with the 6.5m MMT
telescope and the Blue Channel spectrograph.  Observations occurred during the
course of our hypervelocity star survey program on the nights of 2005 Dec 5-6, 2006
May 24-25, 2006 June 20, 2006 Dec 27, and 2007 Mar 18.  We operated the Blue Channel
spectrograph with the 832 line mm$^{-1}$ grating in 2nd order and with a
1.25$\arcsec$ slit.  These settings provided a wavelength coverage of 3650 \AA\ to
4500 \AA\ and a spectral resolution of 1.2 \AA.  One object (an H {\sc ii} region)
was re-observed with the 300 line mm$^{-1}$ grating and a 1 $\arcsec$ slit,
providing wavelength coverage from 3400 \AA\ to 8600 \AA\ with a spectral resolution
of 6.2 \AA.  Exposure times were 30 minutes.  We obtained comparison lamp exposures
after every exposure.  The wavelength solutions are determined from 44 lines with
typical root-mean-square residuals of $\pm0.05$ \AA, or $\pm4$ km s$^{-1}$.  We note
that the single slit spectrograph is a compact instrument with minimal flexure:  
wavelength solutions shift by less than 1 pixel (0.355 \AA) during a night, easily
measured from individual comparison lamp exposures.

\subsection{Spectroscopic Identifications}

	Ten blue plume objects are stars of B spectral type and two are H {\sc ii}
regions.  Figure \ref{fig:spectra} plots the spectra of the ten stars and the two H
{\sc ii} regions, summed and shifted to the rest frame.  The signal-to-noise ratios
($S/N$) of the individual spectra range from $S/N=6$ to 15 per pixel at 4000 \AA,
and depend on target's apparent magnitude and the seeing conditions of the observation.

	The ten B-type stars have visibly narrower Balmer lines and thus lower 
surface gravity than the other B-type stars in the \citet{brown06b, brown07a}
hypervelocity star survey.  Cross-correlation with MK spectral standards
\citep{gray03} indicates that the stars are probably luminosity class I or II B
supergiants, consistent with the stars' inferred luminosities.

	At the distance modulus of Leo A $(m-M)_0 = 24.51 \pm 0.12$
\citep{dolphin02}, the ten B-type stars have absolute magnitudes ranging from $M_V =
-5.3$ to $-3.4$.  For comparison, \citet{corbally84} give absolute magnitudes
$M_V=-5.5$ for a B9 Ib star and $M_V=-3.1$ for a B9 II star.  We conclude the ten
stars are likely B supergiants in Leo A.  Such B supergiants have ages ranging from
$\sim$30 Myr for the most luminous stars to $\sim$200 Myr for the least luminous
stars \citep{schaller92}.

\subsection{Radial Velocities}

	We measure radial velocities with the cross-correlation package RVSAO
\citep{kurtz98}.  We begin by observing the B9 II star $\gamma$ Lyr by quickly
scanning the star across the spectrograph slit.  This procedure provides us with a
very high signal-to-noise ratio cross-correlation template with a known velocity
\citep{evans67, gray03}.  The accuracy of the velocity zero-point comes from the
error on the mean of the 44 comparison lamp lines used to determine the template's
wavelength solution, $\pm0.6$ km s$^{-1}$.

\begin{figure}		
 \plotone{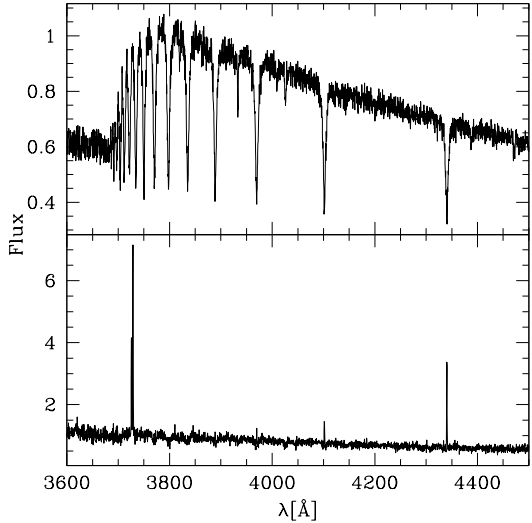}
 \caption{ \label{fig:spectra}
	MMT spectra of the ten B supergiants ({\it upper panel}) and the two H {\sc 
ii} regions ({\it lower panel}), summed together and shifted to rest frame.  The
continuum fluxes are arbitrarily normalized. }
 \end{figure}

	Is is important that we maximize velocity precision for our velocity
dispersion measurement, and we achieve the best precision by cross-correlating the
stars with themselves.  Thus, after measuring the stars' velocities with the
$\gamma$ Lyr template, we shift the spectra to the rest frame and sum them together
to create a second template (shown in Figure \ref{fig:spectra}).  We then
cross-correlate the ten stars with this second template of themselves.  Table
\ref{tab:obs} lists the resulting heliocentric radial velocities and errors.  The
mean cross-correlation precision is $\pm3.7$ km s$^{-1}$.

	We also measure the radial velocities of the H {\sc ii} regions with RVSAO,
but this time using Gaussian fits to the emission lines.  The final velocity of SDSS
J095927.532+304457.75 comes from a weighted mean of the 3727 [O{\sc ii}] doublet
(resolved in our spectra), H$\delta$, and H$\gamma$ emission lines.  A
low-dispersion spectrum of SDSS J095933.320+304439.21 provides additional line
measurements from H$\beta$, [O{\sc iii}], and H$\alpha$ for that object.  The
velocity of SDSS J095933.320+304439.21 is the weighted average of all of its
observed lines. The mean emission-line velocity error is $\pm3.9$ km s$^{-1}$.

\section{RESULTS}

\subsection{Stellar Velocity Dispersion}

\begin{figure}		
 \plotone{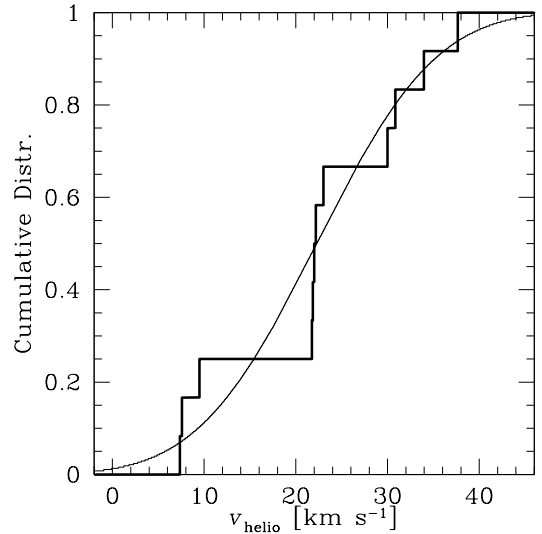}
 \caption{ \label{fig:vel}
	Cumulative distribution of the observed velocities ({\it histogram})  
compared to a Gaussian distribution ({\it curve}) with dispersion 10.0 km s$^{-1}$
and mean velocity 22.3 km s$^{-1}$.}
 \end{figure}

	The average velocity of our twelve Leo A objects is $22.3 \pm 2.9$ km
s$^{-1}$ (see Figure \ref{fig:vel}), statistically identical with the $23 \pm 3$ km
s$^{-1}$ systemic H {\sc i} velocity measured by \citet{allsopp78} and the 23.2 -
24.0 km s$^{-1}$ systemic H {\sc i} velocities measured by \citet{young96}.  Thus
the velocities of our twelve objects are all consistent with membership in Leo A.

	The root-mean-square velocity dispersion of our twelve objects is 10.0 km
s$^{-1}$.  We derive the intrinsic velocity dispersion by subtracting in quadrature
the average 3.8 km s$^{-1}$ uncertainty of the observations.  Thus we measure an
intrinsic stellar velocity dispersion of $\sigma = 9.3 \pm 1.3$ km s$^{-1}$.  

	We estimate the robustness of the velocity dispersion measurement by
comparing the cumulative distribution of velocities to a Gaussian distribution (see
Figure \ref{fig:vel}).  A Kolmogorov-Smirnov test finds a 0.5 likelihood of drawing
the twelve objects from a Gaussian distribution with the observed velocity
dispersion.  Greater number statistics are always desirable, but it appears that the
twelve blue plume objects provide a statistically sound measurement of Leo A's
stellar velocity dispersion.

	Our stellar velocity dispersion measurement is identical to the H {\sc i} 
gas velocity dispersion measured by \citet{young96}:  $9.3 \pm 1.4$ km s$^{-1}$.  
\citet{young96} also observe an H {\sc i} component with $3.5 \pm 1.0$ km s$^{-1}$
dispersion localized in high column-density regions.  If we remove the two H {\sc
ii} regions from our own analysis, the B-type stars have a mean velocity of $21.5
\pm 3.4$ km s$^{-1}$ and an intrinsic velocity dispersion of $\sigma_{B} = 10.1 \pm
1.3$ km s$^{-1}$.  This dispersion is statistically identical to our original value.  

	There is no evidence for rotation of the stellar component of Leo A; the
high- and low-velocity blue plume objects appear inter-mixed on the sky.  This
result is consistent with absence of rotation seen in the H {\sc i} gas \citep{lo93,
young96}.  Given that detailed H {\sc i} maps show velocity structure in Leo A
\citep{young96}, it is possible that additional observations may reveal structure in
the stellar radial velocity distribution.

\begin{deluxetable*}{lrcccccl}           
\tablewidth{0pt}
\tablecaption{LEO A BLUE PLUME OBJECTS\label{tab:obs}}
\tablecolumns{7}
\tablehead{
  \colhead{RA} & \colhead{Dec} & \colhead{type} & \colhead{$v_{helio}$} &
  \colhead{$g'$} & \colhead{$(u'-g')_0$} & \colhead{$(g'-r')_0$} \\
  \colhead{J2000} & \colhead{J2000} & \colhead{} & \colhead{{\small km s$^{-1}$}} &
  \colhead{mag} & \colhead{mag} & \colhead{mag}
}
        \startdata
9:59:15.124 & 30:44:10.40 & B         & $23.0 \pm 2.5$ & 19.896 &  0.761 & -0.279 \\
9:59:16.940 & 30:43:48.22 & B         & $21.7 \pm 5.2$ & 19.050 & -0.021 & -0.353 \\
9:59:20.223 & 30:43:52.71 & B         & $34.0 \pm 2.8$ & 19.435 &  0.458 & -0.375 \\
9:59:23.220 & 30:45:06.23 & B         & $ 7.4 \pm 3.8$ & 20.026 &  0.584 & -0.347 \\
9:59:24.909 & 30:44:36.69 & B         & $30.9 \pm 3.4$ & 19.797 &  1.004 & -0.257 \\
9:59:25.980 & 30:46:10.44 & B         & $ 9.5 \pm 5.0$ & 20.964 &  0.652 & -0.397 \\
9:59:26.351 & 30:45:26.09 & B         & $37.6 \pm 2.3$ & 19.131 &  0.412 & -0.300 \\
9:59:27.058 & 30:45:38.79 & B         & $21.9 \pm 4.7$ & 20.267 &  0.982 & -0.388 \\
9:59:27.326 & 30:45:44.69 & B         & $ 7.6 \pm 4.9$ & 20.661 &  1.082 & -0.420 \\
9:59:27.532 & 30:44:57.75 & H{\sc ii} & $30.0 \pm 3.6$ & 19.984 & -0.238 & -0.607 \\
9:59:32.129 & 30:43:48.55 & B         & $22.0 \pm 2.6$ & 20.471 &  0.383 & -0.360 \\
9:59:33.320 & 30:44:39.21 & H{\sc ii} & $22.2 \pm 4.2$ & 19.520 & -0.288 & -0.252 \\
        \enddata
 \end{deluxetable*}

\subsection{Mass-to-Light Ratio}

	We now estimate the kinematic mass of Leo A.  Because there is no evidence
for rotation, we assume that the galaxy is in pressure equilibrium and apply two
simple mass estimators:  the virial theorem, and the projected mass estimator of
\citet{heisler85}.  The virial mass is given by $M_{vir} = 696 R_e \sigma_z^2 ~
M_{\sun}$, where $R_e$ is the effective radius in pc and $\sigma_z$ is the
one-dimensional velocity dispersion in km s$^{-1}$.  Our objects are located inside
a radius of $2\arcmin = 500$ pc, while Leo A's observed stellar distribution extends
to a radius of $8\arcmin = 2000$ pc.  If we choose $R_e=500$ pc, Leo A's virial mass
is $M_{vir}\sim3\times10^7 ~M_{\sun}$.

	The virial theorem, however, is both more biased and less stable for small
numbers of test particles than is the projected mass estimator \citep{bahcall81}. 
Thus we use the \citet{heisler85} projected mass estimator to obtain a more accurate
estimate of Leo A's mass:
	\begin{equation}
M_{proj} = \frac{f}{G(N-\alpha)} \sum_{i=1}^{N} V_{z,i}^2 R_{\bot,i}
	\end{equation} where $G$ is the gravitational constant, $N$ is the number of
stars, $\alpha$ is an empirical correction to the center of mass (Heisler et al.\ 
use $\alpha=1.5$), $V_z$ is the velocity relative to the mean, $R_\bot$ is the 
projected separation from the center of the galaxy, and $f$ is a constant that 
depends on the eccentricity of the stellar orbits.  For purely isotropic orbits  
$f=32/\pi$, while for purely radial orbits$f=64/\pi$.
	Using the velocities and positions in Table \ref{tab:obs}, Leo A has a
kinematic mass of $5.3\pm1.3\times10^7 ~M_{\sun}$ for purely isotropic orbits and
$10.6\pm2.6\times10^7 ~M_{\sun}$ for purely radial orbits.  Heisler et al.\ prefer
using the smaller mass derived from isotropic orbits, but for purposes of
discussion, we will assume that Leo A's mass is the average of the two projected
mass estimates:  $8\times10^7~M_{\sun}$.
	This mass is derived from objects inside a radius of $2\arcmin = 500$ pc.

	By comparison, \citet{vansevicius04} estimate that Leo A's stellar mass is
$M_{stars}=4\pm2\times10^6$ M$_{\sun}$, consistent with the galaxy's optical
luminosity.  More recently, \citet{lee06} use {\it Spitzer} 4.5\micron\ imaging to
estimate that Leo A's total stellar mass is $M_{stars}=0.8\times10^6$ M$_{\sun}$
with an uncertainty of 0.5 dex.  These stellar mass estimates are factors of 20 - 
100 times smaller than our kinematic mass estimate.

	Leo A's total $V$-band luminosity is $M_V=-11.7$, which comes from the
apparent magnitude $V_{tot} =12.8 \pm 0.2$ \citep{mateo98} and the distance modulus
$(m-M)_0 = 24.51 \pm 0.12$ \citep{dolphin02}.  Assuming the Sun has $M_{V, \sun} =
+4.8$, Leo A's total luminosity in solar units is $4\times10^6 ~L_{\sun}$.
	Thus the mass-to-light ratio of Leo A is $M/L_{tot} = 20\pm6
~M_{\sun}/L_{\sun}$ for a mass of $8\times10^7 ~M_{\sun}$.  Because our
spectroscopic targets do not sample the full extent of Leo A, this mass-to-light
ratio is a lower limit to Leo A's true mass-to-light ratio.

	A mass-to-light ratio of 20 suggests that Leo A is dominated by dark matter.  
\citet{young96} reach the opposite conclusion from their H {\sc i} velocity
dispersion, but their result is explained by the revision of Leo A's distance from
2.2 Mpc to 800 kpc.  Leo A's total H {\sc i} mass within $a=8\arcmin$ is
$M_{HI}=1.0\pm0.2\times10^7 ~M_{\sun}$ \citep{allsopp78, young96} at a distance of
800 kpc.  The gas mass includes the 10\% correction for helium gas.  Thus baryonic
matter -- stars plus gas -- accounts for at most $\sim$20\% of Leo A's total mass.

\section{DISCUSSION AND CONCLUSIONS}

	We have obtained spectroscopy for twelve blue plume objects in the central
$2\arcmin$ of Leo A.  Ten of these objects are young B supergiants.  We measure a
stellar velocity dispersion of $\sigma = 9.3 \pm 1.3$ km s$^{-1}$, identical to Leo 
A's H {\sc i} gas dispersion \citep{young96}.  From this we estimate a projected 
mass of $8\pm2.7\times10^7 ~M_{\sun}$, which implies that Leo A's mass is at least 
$\sim$80\% dark matter.

	Dwarf galaxies are thought to be the smallest bodies containing dynamically
significant amounts of dark matter, and so it is interesting to place Leo A in the
context of cosmological simulations.  \citet{evrard07} show that the velocity
dispersion of dark matter halos follow a tight correlation with total mass,
$\sigma_{\rm DM} = (1084\pm13 ~{\rm km s^{-1}}) (h(z) M_{200}/10^{15}
~M_{\sun})^{0.3359\pm0.0045}$, where $M_{200}$ is the mass within a sphere with mean
interior density 200 times the critical density.  Leo A's mass, $8\times10^7
~M_{\sun}$, would fill such a sphere with a radius of $r_{200}=9$ kpc.  The halo
virial relation is derived from $\sim10^{15} ~M_{\sun}$ dark matter halos, but
Evrard et al.\ show it is valid down to $\sim10^{10} ~M_{\sun}$ halos.  If we simply
equate Leo A's mass to $M_{200}$, the halo virial relation predicts $\sigma_{DM}=4$
km s$^{-1}$ for $h(z)=0.70$.  This prediction is less than half of the observed
velocity dispersion.  One possible explanation for the discrepancy is that Leo A has
not reached dynamical equilibrium, and thus its velocity dispersion is inflated
\citep{young96}.  Or, perhaps the discrepancy suggests that dwarfs like Leo A
experience a different evolutionary path than a purely hierarchical growth of dark
matter halos.

	Remarkably, Leo A's stellar velocity dispersion is very similar to that of
Local Group dwarf spheroidals (dSphs), which have central velocity dispersions of 8
to 10 km s$^{-1}$ \citep{mateo98}.  One explanation for the common central velocity
dispersion is that all Local Group dwarfs are enclosed in dark matter halos of
similar total mass \citep{mateo93}. Galaxies with smaller velocity dispersions
(total mass $\lesssim10^8 ~M_{\sun}$) are possibly re-ionized and thus never form
stars \citep[e.g.][]{navarro97}.  If this picture is correct, then the total mass to
light ratio of a dwarf is a function of its luminosity $(M/L)_{tot} = M_{DM}/L +
(M/L)_*$, where $M_{DM}$ is the fixed dark matter halo mass, $L$ is the total
$V$-band luminosity, and $(M/L)_*$ is the stellar mass to light ratio.

\begin{figure}		
 \plotone{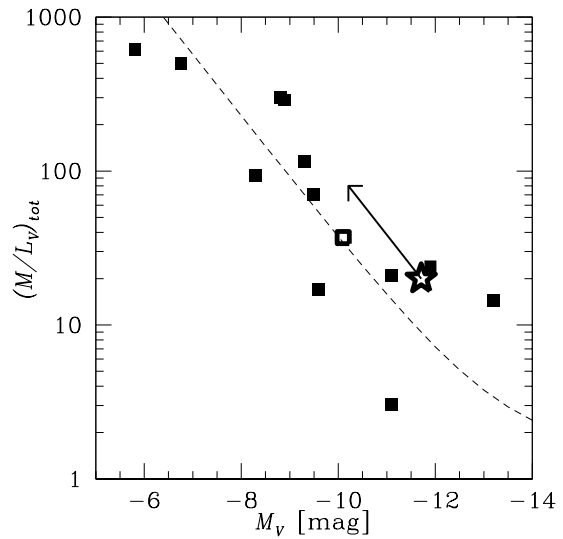}
 \caption{ \label{fig:ml}
	Mass-to-light ratio of Local Group dSph galaxies with masses determined from
central velocity dispersions ({\it solid squares}), adapted from \citet{mateo98} and
\citet{koch07}.  The dashed line is the $(M/L)_{tot}$ relation for a galaxy in a
dark matter halo of constant mass $3\times10^7 M_{\sun}$.  We estimate $M/L=20\pm6$
for Leo A ({\it star}), which falls near the fixed halo mass relation.  The arrow
indicates what happens if Leo A stops forming stars and fades to a dSph-like color.  
The Phoenix transition dwarf ({\it open square}) also agrees with the fixed halo
mass relation.}
 \end{figure}

	In Figure \ref{fig:ml} we plot the $(M/L)_{tot}$ versus $V$-band luminosity
for Local Group dSphs with central velocity dispersion measurements.  We note that
Leo A has a central velocity dispersion and no observed rotation, thus its dynamical
mass is directly comparable with dSphs.  Dwarf irregulars have masses determined
from rotation and are not directly comparable.  We base Figure \ref{fig:ml} on the
\citet{koch07} version of \citet{mateo98}'s plot.  The solid squares are And II
\citep{cote99}, And IX \citep{chapman05}, Bo\"otes \citep{belokurov06a, munoz06a},
Carina and Sextans \citep{wilkinson06}, Draco and Ursa Minor \citep{wilkinson04},
Fornax \citep{wang05}, Leo I \citep{koch07}, Leo II and Sculptor \citep{mateo98},
and Ursa Major \citep{willman05, kleyna05}.  The dashed line shows the $(M/L)_{tot}$
relation for a fixed dark matter halo mass $M_{DM}=3\times10^7 ~M_{\sun}$ and
stellar $(M/L)_*=1.5 ~M_{\sun}/L_{\sun}$ \citep{koch07}.  Leo A, plotted as a star,
falls very near the fixed halo mass relation for dSphs.

	However, Leo A's stellar population is quite different from that of the
dSphs.  Integrated colors provide a quantitative measure of the difference:  Leo A
has $(\bv)=0.15$, systematically bluer than the average dSph with $(\bv)=0.8\pm0.25$
\citep{mateo98}.  As its stellar population ages, Leo A's luminosity will decrease
and its mass to light ratio will increase.  We estimate this change using
Starburst99 \citep{leitherer99, vazquez05} with $Z=0.0004$ Padova tracks.  We find
that in a couple of Gyr, assuming Leo A has no further star formation, it will reach
$(\bv)=0.8$ and will have faded $\sim1.5$ magnitudes in $M_V$.  We indicate this
evolution with the arrow in Figure \ref{fig:ml}.  Leo A still falls well within the
observed scatter around the $(M/L)_{tot}$ relation.

	Comparing Leo A with ``transition dwarfs'' may be more fair than comparing
with dSphs.  Transition dwarfs have old stellar populations like dSphs, but also
contain gas and young stars like Leo A.
	A central velocity dispersion is available for the Phoenix transition dwarf
\citep{mateo98} (the open square in Figure \ref{fig:ml}), which places it squarely
on the $(M/L)_{tot}$ relation.  Thus, despite their different star formation
histories, Leo A, Phoenix, and the dSphs appear to share remarkably similar
kinematics and dark matter halo mass.

	If transition dwarfs represent the stage between gas-rich dwarf irregulars
and gas-poor dSphs, this evolution must involve some amount of galaxy interaction.
	Most dSphs in the Local Group are located near the major spirals, so the
dSphs' lack of gas and young stars likely results from repeated gravitational and/or
hydrodynamic interactions with the spirals.
	In a comprehensive study of minor galaxy interactions in the SDSS,
\citet{freedman07} find that the lowest luminosity galaxies in close pairs
experience the largest fractional boosts in their specific star formation rates.  
Perhaps Leo A's episodic star formation history is a history of its interactions 
with objects in the Local Group.

	One clue to the evolution of transition dwarfs in the Local Group may come
from comparison of the rotation velocity and central velocity dispersion.  A wide
variety of studies demonstrate the relation between these kinematic measures and the
formation history of galaxies \citep[e.g.][]{pizzella05, jesseit05, derijcke05,
derijcke06}.  Multi-slit spectrographs can now provide radial velocities for 
hundreds of stars in nearby dwarfs, making such studies possible for the first time.

\acknowledgements

	We thank K.\ Rines for helpful discussions and thank the referee for
comments that improved this paper.
	We thank M.\ Alegria, J.\ McAfee, and A.\ Milone for their assistance with
observations obtained at the MMT Observatory, a joint facility of the Smithsonian
Institution and the University of Arizona.
	This project made use of data products from the Sloan Digital Sky Survey,
which is managed by the Astrophysical Research Consortium for the Participating
Institutions.
	This research made use of the Smithsonian/NASA Astrophysics Data System
Bibliographic Services.
	This work was supported in part by W.\ Brown's Clay Fellowship and by the
Smithsonian Institution.

{\it Facilities:} \facility{MMT (Blue Channel Spectrograph)}





\end{document}